\documentclass[sigconf,screen=true,anonymous=false,bookmarks=false]{acmart}
\usepackage{lipsum}
\usepackage{titlesec}
\titlespacing\section{0pt}{5pt plus 1pt minus 1pt}{0pt plus 1pt minus 1pt}
\titlespacing\subsection{0pt}{5pt plus 1pt minus 1pt}{0pt plus 1pt minus 1pt}
\titlespacing\subsubsection{0pt}{4pt plus 1pt minus 1pt}{2pt plus 1pt minus 1pt}

\iftrue
\fi

\usepackage{graphicx}
\usepackage{amsmath}

\usepackage{mathtools}
\usepackage{comment}
\usepackage[subrefformat=parens,labelformat=parens]{subfig}
\usepackage{bm}
\usepackage{multirow}
\usepackage{threeparttable,booktabs}
\usepackage{blkarray}
\usepackage{tikz}
\usepackage{balance}
\usepackage{courier}                                       
\usepackage{cleveref}                                      
\usepackage[mathcal]{eucal}
\usepackage[]{algpseudocode}                               
\algrenewcommand\textproc{\texttt}
\makeatletter\let\float@addtolists\relax\makeatother
\usepackage{algorithm}
\usepackage{filecontents}                                  
\usepackage{pgfplots}
\usepackage{pgfplotstable}
\pgfplotsset{compat=newest}
\usepackage[figuresright]{rotating}

\renewcommand{\vec}[1]{\boldsymbol{#1}}

\theoremstyle{plain}

\theoremstyle{definition}
\newtheorem{mydefinition}{\textbf{Definition}}
\newtheorem{myproblem}{\textbf{Problem}}

\definecolor{myblue}{RGB}{154, 164, 208}
\definecolor{myred}{RGB}{242, 200, 195}

\copyrightyear{2020}
\acmYear{2020}
\setcopyright{acmcopyright}\acmConference[ICCAD '20]{IEEE/ACM International Conference on Computer-Aided Design}{November 2--5, 2020}{Virtual Event, USA}
\acmBooktitle{IEEE/ACM International Conference on Computer-Aided Design (ICCAD '20), November 2--5, 2020, Virtual Event, USA}
\acmPrice{15.00}
\acmDOI{10.1145/3400302.3415705}
\acmISBN{978-1-4503-8026-3/20/11}

\begin{document}

\title{
  DAMO: Deep Agile Mask Optimization for Full Chip Scale
}

\author{Guojin Chen}
\affiliation{
    \institution{Chinese University of Hong Kong}
}
\email{cgjhaha@qq.com}
\author{Wanli Chen}
\affiliation{
    \institution{Chinese University of Hong Kong}
}
\email{wlchen@cse.cuhk.edu.hk}
\author{Yuzhe Ma}
\affiliation{
    \institution{Chinese University of Hong Kong}
}
\email{yzma@cse.cuhk.edu.hk}
\author{Haoyu Yang}
\affiliation{
    \institution{Chinese University of Hong Kong}
}
\email{hyyang@cse.cuhk.edu.hk}
\author{Bei Yu}
\affiliation{
    \institution{Chinese University of Hong Kong}
}
\email{byu@cse.cuhk.edu.hk}

\begin{abstract}
    Continuous scaling of the VLSI system leaves a great challenge on manufacturing,
    thus optical proximity correction (OPC) is widely applied in conventional design flow for manufacturability optimization. 
    Traditional techniques conduct OPC by leveraging a lithography model but may suffer from prohibitive computational overhead.
    In addition, most of them focus on optimizing a single and local clip instead of addressing how to tackle the full-chip scale. 
    In this paper, we present DAMO, a high performance and scalable deep learning-enabled OPC system for full-chip scale.
    It is an end-to-end mask optimization paradigm that contains a deep lithography simulator (DLS) for lithography modeling and a deep mask generator (DMG) for mask pattern generation. 
    Moreover, a novel layout splitting algorithm customized for DAMO is proposed to handle full-chip OPC problem.
    Extensive experiments show that DAMO outperforms state-of-the-art OPC solutions in both academia and industrial commercial toolkit.
\end{abstract}

\begingroup
\def\UrlFont{\ttfamily\normalsize}
\mathchardef\UrlBreakPenalty=10000
\maketitle
\endgroup

\section{Introduction}
\label{sec:intro}

Continuously shrinking down of the VLSI system has brought inevitable lithograph proximity effects and hence results in a degradation on manufacturing yield \cite{DFM-TCAD2013-Pan}.
Optical proximity correction (OPC) compensates lithography proximity effects by adding assistant features and moving design edge segments inward or outward \cite{OPC-ASPDAC2020-Yang}.
Mainstream OPC solutions include rule-based OPC \cite{park2000efficient}, model-based OPC \cite{OPC-DATE2015-Kuang,OPC-TCAD2016-Su,OPC-JM3-2016-Matsunawa}, inverse lithography technique (ILT)-based OPC \cite{OPC-DAC2014-Gao,OPC-ICCAD2017-Ma},
and machine/deep learning-based OPC \cite{OPC-DAC2018-Yang,OPC-ASPDAC2019-Jiang,OPC-TCAD2020-Geng}.

Kuang \textit{et al.}~\cite{OPC-DATE2015-Kuang} presented a model-based OPC for faster convergence and achieved good EPE with minor PV Band overhead using multi-stage SRAF insertion and OPC.
Gao \textit{et al.}~\cite{OPC-DAC2014-Gao} tackled the mask optimization problem by solving an ILT formulation,
which descends the gradient of wafer-target error over input masks.
The pixel-based optimization of ILT solution makes them robust to process variations.
The generality of ILT also enables simultaneous mask optimization and layout decomposition as introduced in \cite{OPC-ICCAD2017-Ma,OPC-DAC2020-Zhong}.
These methods, to some extent, improve OPC from quality, robustness, and efficiency. 

The great development of machine learning algorithms has demonstrated the potential of applying artificial intelligence to benefit modern OPC flows.
On one hand, machine learning-guided mask optimization targets to directly generate masks that are close to an optimal status and only fewer fine-tune steps using traditional OPC engines are required to obtain the final mask.
Yang \textit{et al.}~\cite{OPC-DAC2018-Yang} proposed GAN-OPC which grasps the advantage of generative machine learning models that can learn a design-to-mask mapping and provides better initialization of the ILT engine.
On the other hand, machine learning-based lithography simulation aims to speed-up OPC flows by replacing costly lithography simulation with efficient learning models.
Jiang \textit{et al.}~\cite{OPC-ASPDAC2019-Jiang} applied an XGBoost \cite{chen2016xgboost} learning model to predict EPE at certain OPC control points that can guide the adjustment of shape edges.
Instead of predicting wafer image errors, Ye \textit{et al.}~\cite{DFM-DAC2019-Ye}  proposed LithoGAN to build a generative learning model that directly predicts lithography contours.
However, LithoGAN only targets a single shape within a clip, which strictly limits its usage in general OPC tasks.

\begin{figure}[tb!]
    \includegraphics[width=0.98\linewidth]{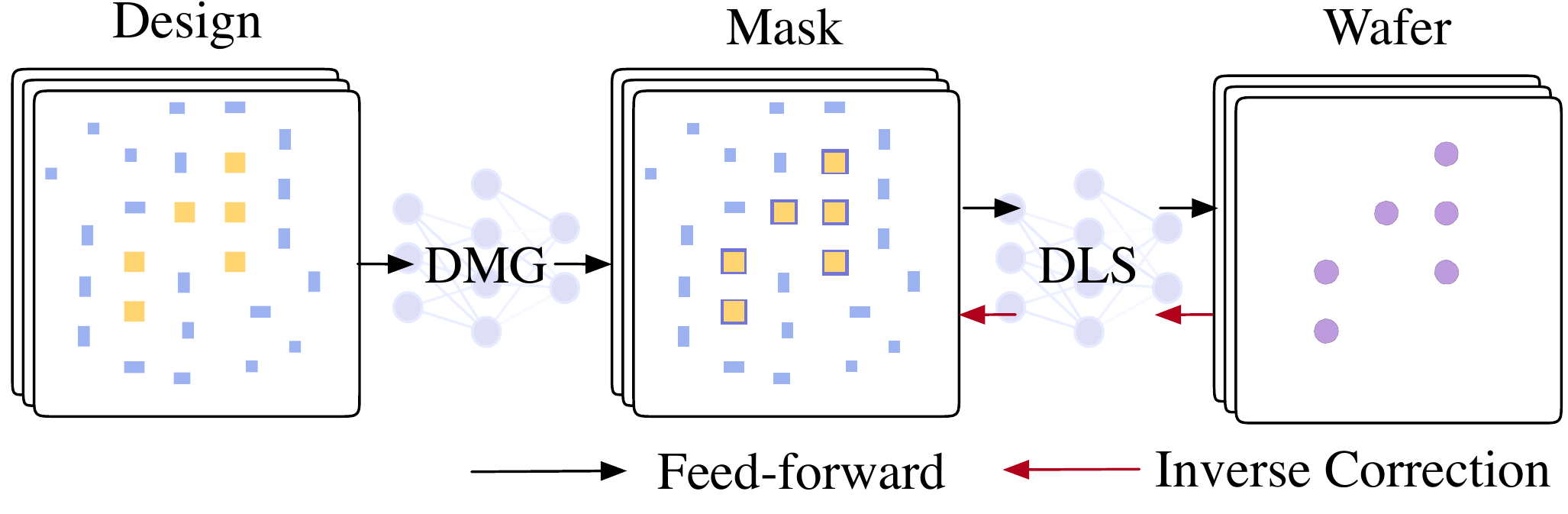}
    \centering
    \caption{
        Overview of our proposed DAMO framework, which consists of two deep networks: deep mask generator (DMG) and deep lithography simulator (DLS).
        The OPC process is completed by utilizing the inverse correction gradient back-propagated from the DLS stage (red arrows).
    }
    \label{fig:overview} 
\end{figure}

There are several issues in previous methods.
Firstly, the model-based/ILT inevitably methods require massive calls of the costly lithography simulation and the mask optimization, both of which are time-consuming.
Secondly, all the previous works in machine learning-guided OPC limit the single-clip input layout into a low-resolution such as $256 \times 256$ pixel image.
They are all exhibiting drawbacks that have still to go through traditional OPC engines in final steps due to the low-resolution limits.
Since the resolution loss is intolerable in OPC, the usage scenarios of previous work in machine learning-guide OPC are limited.
And worse still, the machine learning-based single-clip OPC is not practical.
Thirdly, despite a variety of methods have been proposed, most of them focused on how to optimize a given single clip, and rarely discussed how to tackle the OPC problem in a view of the full-chip scale. 
For full-chip OPC tasks, the biggest barrier to conventional methods is the runtime overhead.
Pang \textit{et al.}~\cite{pang2019study} presented D2S to create full-chip ILT in a single day with giant GPU/CPU pairs, which consumes a large amount of resources on the handcrafted hardware and software. 
The learning-based methods, to the best of our knowledge, have not achieved any progress on full-chip mask optimization due to the dataset limitation and the low wafer pattern fidelity.


To address these concerns, we present DAMO, a unified OPC engine that is equipped with high-resolution GANs for full-chip scale.
Deep convolution GANs (DCGAN) \cite{GAN-ICLR2016-DCGAN} has been demonstrated to be successful in generating high-resolution images. 
In DAMO, we designed DCGAN-HD which is customized from DCGAN with a high-resolution generator and multi-scale discriminators with perceptual losses.
Then we design a deep lithography simulator (DLS) based on DCGAN-HD that takes the input of mask and generates the lithography contours faster
with similar contour quality compared to legacy lithography simulation process.
The DLS design also enables a unified neural network-based OPC framework where another deep mask generator (DMG) engine is trained along with the gradient back-propagated from DLS, which allows direct output of optimized high-quality masks (as shown in \Cref{fig:overview}).
We further propose a stitchless full-chip splitting algorithm, with which we can perform full-chip OPC tasks efficiently with a few GPU resources.
Our main contributions are as follows:
\begin{itemize}
    \item We design DCGAN-HD, a very competitive high-resolution feature extractor ($1024 \times 1024$) by redesign the generator and discriminator of DCGAN.
    \item We build up DLS and DMG based on DCGAN-HD. DLS is expected to conduct high-resolution lithography simulation. By training along with the inverse correction from DLS, DMG can directly generate high-quality masks.
    \item We develop an efficient stitchless full-chip splitting algorithm to apply DAMO on a layout of any size.
    \item We compare our proposed framework with state-of-the-art commercial tool Calibre \cite{TOOL-calibre}: $5\times$ speed-up in single-clip OPC tasks and $1.3\times$ acceleration in full-chip OPC tasks, while maintaining an even better solution quality.
\end{itemize}

The rest of the paper is organized as follows:
\Cref{sec:prelim} introduces terminologies and evaluation metrics related to this work.
\Cref{sec:algo} details the proposed DAMO architecture.
\Cref{sec:train} shows the data preparation and DAMO training procedure, while \Cref{sec:split_algo} provides the full-chip splitting algorithm.
\Cref{sec:result} details experimental results and followed by conclusion in \Cref{sec:conclu}.

\section{Preliminaries}
\label{sec:prelim}

In this section, we will introduce some concepts and background related to this work and the problem formulation.
\subsection{cGAN Basis}
cGAN is the short for conditional Generative Adversarial Networks \cite{pixel2pixel,cGAN},
which resembles classical GANs \cite{GAN-NIPS2014-Ian} that consists of a generator and a discriminator.
The generator is trained to generate patterns follow some distribution such that the discriminator cannot identify whether these data comes from the generator or the training dataset.
cGAN differs from GANs by certain constraints such that inputs and outputs of the generator can have stronger beneath connections.
Representative cGAN applications in VLSI include GAN-OPC \cite{OPC-DAC2018-Yang} and LithoGAN \cite{DFM-DAC2019-Ye}.
The former is designed for layout mask synthesis and the latter focuses on lithography contour prediction of the single via/contact shapes.

\subsection{Problem Formulation}
We introduce the following terms and evaluation metrics for the DAMO framework.
\begin{mydefinition}[mIoU]
Given two shapes $P$ and $G$, the IoU between $P$ and $G$ is $ IoU(P,G) = {P \cap G} / {P \cup G}$. The mIoU is mean IoU.
\end{mydefinition}
\begin{mydefinition}[Pixel Accuracy]
Pixel accuracy (pixAcc) is defined as the percentage of pixels that are correctly classified on an image.
\end{mydefinition}
Additionally, we have two evaluation metrics to measure mask quality following \cite{OPC-DAC2018-Yang}.
The squared $L_2$ error measures the quality of a mask under nominal process conditions, while PV Band measures the robustness of the generated mask under variations.
\begin{mydefinition}[Squared $L_2$ Error]
	Let $\vec{w}$ and ${\vec{y}}$ as design image and wafer image respectively, the squared $L_2$ error is calculated by $||\vec{w}-\vec{y}||_2^2$.
\end{mydefinition}

\begin{mydefinition}[PV Band]
  Given the lithography simulation contours under a set of process conditions, the PV Bands is
the area among all the contours under these conditions.
  
\end{mydefinition}
With these definitions and evaluation metrics, the problem of mask optimization is defined as follows:
\begin{myproblem}[Mask Optimization]
  Given a design image $\vec{w}$, the objective of mask optimization is generating
  the corresponding mask $\vec{x}$ such that remaining patterns $\vec{y}$ after lithography process is as close as $\vec{w}$
	or, in other words, minimizing PV Band and squared $L_2$ error of lithography images.
\end{myproblem}

\section{DAMO Framework}
\label{sec:algo}

The architecture overview of DAMO is illustrated in \Cref{fig:dcganhd}.
As the first part of DAMO, DLS aims to conduct an efficient and high-quality lithography simulation with the generative neural network model.
Although LithoGAN \cite{DFM-DAC2019-Ye} tries to alleviate the problem by embedding coordinate inputs,
the scenario of application is strictly limited for a single via/contact shape, which is not practical in most cases.
Therefore, DLS is developed as a customized cGAN for general-purpose lithography contour prediction tasks.

DMG is the second part of DAMO, which shares the identical architecture with DLS.
The forward lithography process can be described with the following equation:
\begin{equation}
  \label{eq:forward_litho}
  \vec{Z}=f(\vec{M}).
\end{equation}
The traditional ILT tries to obtain the optimal mask $\vec{M}_{opt}$ based on the given lithography model, which is presented as: 
\begin{equation}
  \label{eq:ilt}
  \vec{M}_{opt}=f^{-1}\left(\vec{Z}_{t}\right),
\end{equation}
where $\vec{Z}_t$ is the design pattern and $\vec{M}_{opt}$ is the optimized mask with OPC.
In DAMO, we regard DLS as $f$ in \Cref{eq:forward_litho}. 
However, different masks may yield the same result, thus \Cref{eq:ilt} is an ill-posed problem.
Previous mask optimizer GAN-OPC \cite{OPC-TCAD2020-Yang} generates masks by using cGAN to learn the mapping between the design and the mask pattern.
Inspired from conventional ILT, our DMG steps further by not only learning mask patterns from training datasets but also being optimized by gradient back-propagated from the pre-trained DLS.
After training, the generator of DMG performs inference to generate the solutions. 


\subsection{Improving Accuracy by Higher Resolution}

Different from synthesizing photo-realistic images in computer vision tasks, the OPC task using generative models has its own properties.
Intuitively, the layout in the OPC task has simpler patterns (mostly rectangles) but higher precision demands compared with image translation tasks. 
Moreover, the inputs of traditional image generation tasks are fixed-size images whose width or height is barely more than 2048 pixels. However, layouts contain thousands of via/contacts or SRAF patterns, 
whose area can reach more than 100$\times$100 $um^2$. 
Previous work GAN-OPC \cite{OPC-DAC2018-Yang} converts 1000$\times$1000 $nm^2$ layout to 256$\times$256 pixel images, which means 1-pixel shift error will cause an 8 nm shift in the output layout, making the results vulnerable for the industrial OPC tasks.
To eliminate image transformation error, we set the input resolution of our model to be 1024$\times$1024 pixels to contain the full 1024$\times$1024 $nm^2$ layout.
Combined with the window splitting algorithm which will be introduced in \Cref{sec:result}, DAMO framework can process input layout of any size, even the large full-chip layouts.

It is known that the adversarial training might be unstable and hard to converge for high-resolution image generation tasks, as mentioned in \cite{Chen_2017_ICCV,GAN-ICLR2016-DCGAN,wang2018high}. 
Therefore, we present DCGAN-HD, a new conditional GANs model qualified with high-resolution input images, which is the basic architecture of DLS and DMG.


\subsection{DCGAN-HD: Solution for High Resolution}
Previous work GAN-OPC is a conditional GAN framework for design to mask translation which consists of a generator $G$
and a discriminator $D$. It adopts U-Net \cite{U-Net} as the generator with the input resolution of 256$\times$256,
We tested the GAN-OPC framework directly on high-resolution images and found the training is unstable and the generated results usually became empty.
DCGAN \cite{GAN-ICLR2016-DCGAN} is one of the popular and successful network designs for cGAN allowing for higher resolution and deeper models.
Based on DCGAN we present DCGAN-HD, a robust high-resolution conditional GAN model consisting of a newly designed generator, multi-scale discriminators, and a novel adversarial loss function.
The architecture is illustrated in \Cref{fig:dcganhd}.


\subsubsection{High-resolution Generator for DCGAN-HD.}
\label{subsubsec:h-res-g}

\begin{figure}[tb!]
  \centering
  \includegraphics[width=.98\linewidth]{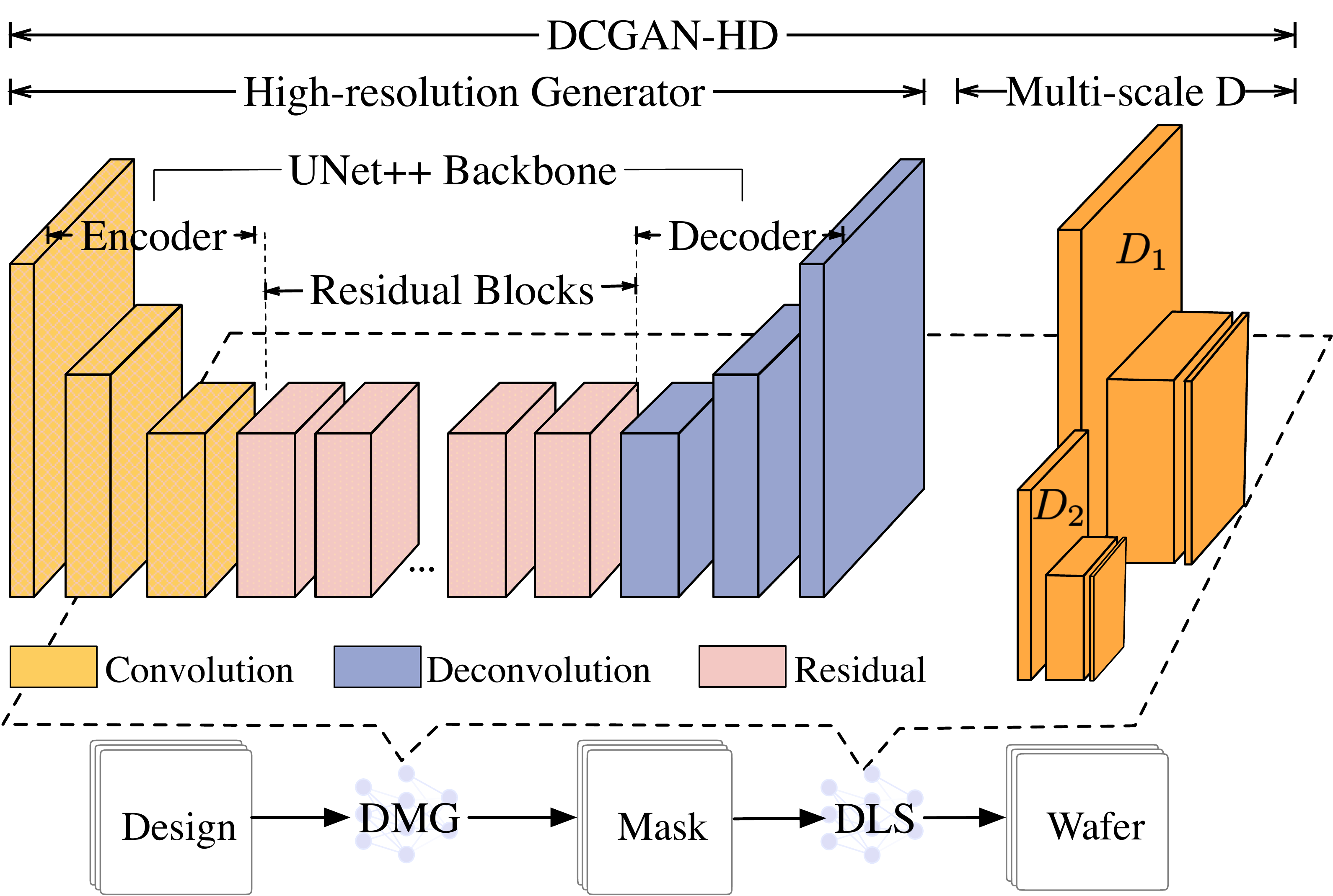}
   \caption{Architecture of DCGAN-HD with high-resolution generator and multi-scale discriminators, used in both DMG and DLS.}
  \label{fig:dcganhd}
\end{figure}

The left part of \Cref{fig:dcganhd} shows the high-resolution generator.
In DLS part, the generator of DCGAN-HD resembles lithography simulation which requires mask-to-wafer mapping.
In DMG part, with the gradient backpropagated from DLS, the generator focus on synthesizing the mask patterns from design and SRAF pattern groups.

\textbf{UNet++ Backbone.} Previous work \cite{OPC-DAC2018-Yang} and \cite{DFM-DAC2019-Ye} adopt traditional UNet \cite{U-Net} for mask generation.
Input features are down-sampled multiple times.
With the decreasing of feature resolution,
it is easier for a network to gather high-level features such as context features
while low-level information such as the position of each shape becomes harder to collect.
However, in OPC tasks, low-level information matters more than in the common computer vision tasks.
For example, the shape and relative distance of design or SRAF patterns must remain unchanged after the deep mask optimization or deep lithography process.
The number and relative distance of via patterns in an input layout have a crucial influence on the result.
The features of OPC datasets determine the vital importance of the low-level features.
UNet++ \cite{zhou2018unet++} is hence proposed for better feature extraction by assembling multiple UNet that have different numbers of downsampling operations.
It redesigns the skip pathways to bridge the semantic gap between the encoder and decoder feature maps,
contributing to the more accurate low-level feature extraction.
The dense skip connections on UNet++ skip pathways improve gradient flow in high-resolution tasks.
Although UNet++ has a better performance than UNet, it is not qualified to be the generator of DCGAN-HD.
For further improvement, we manipulate the UNet++ backbone with the guidelines suggested in DCGAN \cite{GAN-ICLR2016-DCGAN}.
We will show later that our high-resolution generator outperforms UNet and UNet++ by a large margin.


\textbf{Residual blocks.} Most importantly, following Johnson \textit{et al.}~\cite{johnson2016perceptual} settings, a set of residual blocks are added at the bottleneck of UNet++,
which has been proven successful in style transfer and high-resolution image synthesis tasks. Since in OPC tasks, most structures are shared in output and input images (design and SRAFs),
residual connections make it easy for the network to learn the identity function, which is appealing in the mask generation process.
Specifically, we use 9 residual blocks, each of which contains two 3$\times$3 convolution layers and batch normalization layers.

\subsubsection{Multi-scale Discriminators for DCGAN-HD.}
\label{subsubsec:multi-d}
The high-resolution input also imposes a critical challenge to the discriminator design.
A simple discriminator that only has three convolutional layers with LeakyReLU \cite{leakyrelu} and Dropout \cite{JMLR:dropout} is presented.
Since the patterns in OPC datasets have simple and homogeneous distribution, a deeper discriminator has a higher risk of over-fitting.
Therefore, we simplify the discriminator by reducing the depth of the neural network. 
Meanwhile, a dropout layer is attached after each convolutional layer.
We use $3 \times 3$ convolution kernels in generator for parameter-saving purposes and $4 \times 4$ kernels in discriminator to increase receptive fields.

However, during training, we find that the simple discriminator fails to distinguish between the real and the synthesized images when more via patterns occur in a window.
Because when the number of via reaches 5 or 6 in a window, the via patterns will have larger impact on each other and the features become more complicated.
Inspired by Wang \textit{et al.}~in pix2pixHD \cite{wang2018high}, we design multi-scale discriminators.
Different from pix2pixHD \cite{wang2018high} that using three discriminators,
our design uses two discriminators that have an identical network structure but operate at different image scales, which is named $D1$, $D2$, as shown in the right part of \Cref{fig:dcganhd}.
Specifically, the discriminators $D1$, $D2$ are trained to differentiate real and synthesized images at the two different scales, 1024$\times$1024 and 512$\times$512, respectively, which helps the training of the high-resolution model easier. 
In our tasks, the multi-scale design also shows its strengths in flexibility.
For example, when the training set has only one via in a window, we can use only $D1$ to avoid over-fitting and reduce the training time.

\subsubsection{Perceptual Losses.}
\label{subsubsec:per-loss}
Instead of using per-pixel loss such as $L_1$ loss or $L_2$ loss, we adopt
the perceptual loss which has been proven successful in style transfer \cite{johnson2016perceptual},
image super-resolution and high-resolution image synthesis \cite{wang2018high}.
A per-pixel loss function is used as a metric for understanding differences between input and output on a pixel level.
While the function is valuable for understanding interpolation on a pixel level, the process has drawbacks.
For example, as stated in \cite{johnson2016perceptual}, consider two identical images offset from each other by one pixel;
despite their perceptual similarity they would be very different as measured by per-pixel losses.
More than that, previous work \cite{GAN-ICLR2016-DCGAN} shows $L_2$ Loss will cause blur on the output image.
Different from per-pixel loss, perceptual loss function in \Cref{eq:per-loss} compares ground truth image $\vec{x}$ with generated image $\hat{\vec{x}}$ based on high-level representations from pre-trained convolutional neural networks $\Phi$, which is ideal in DAMO framework.
In DLS part, since the wafer pattern is not a regular circle, it is meaningless to fit the exact border of a wafer on the pixel level, the ultimate goal is to generate a better mask with higher perceptual quality wafer,
reflected in less $L_2$ error and smaller PV Band.

\begin{align}
  \label{eq:per-loss}
  \begin{split}
      \mathcal{L}_{L_P}^{G, \Phi}(\vec{x}, \vec{\hat{x}})=& \mathcal{L}_{L_1}({{\Phi}(\vec{x})}, {\Phi(\vec{\hat{x}})})
      = \mathbb{E}_{{\vec{x}}, {\vec{\hat{x}}}}\left[\|{\Phi(\vec{x})}-{\Phi(\vec{\hat{x}})}\|_{1}\right],
  \end{split}
\end{align}

\section{Data preparation and training}
\label{sec:train}

In order to collect sufficient data for training, we develop a data generation pipeline that can generate infinite training data, with which our DCGAN-HD can be fully utilized to simulate the lithography process and generate high-quality mask patterns. 
The overall training procedure of DAMO can be divided into two parts which are depicted in \Cref{fig:wholeflow}.

\subsection{Building Training Set from Scratch}
\label{subsec:build_train}

It takes five steps to generate a training image, including design generation, SRAF insertion (with design rule checking), OPC, lithography simulation and layout to image transformation.

\textbf{Design a design pattern.}
Via patterns are obtained under the following constraints using a layout pattern generator~\cite{yang2019automatic}. 
Firstly, all via patterns (70$\times$70 $nm^2$) are restricted in a 1024$\times$1024 $nm^2$ window.
Secondly, by changing the via density we can control the number of via patterns in a single window. 
The via patterns are grouped evenly by the via numbers for reducing the bias caused by the random distribution of training set.

\textbf{SRAF insertion and DRC.}
Mentor Calibre \cite{TOOL-calibre} is applied to do the SRAF insertion and design rule checking. 
Since the design area is 1024$\times$1024 $nm^2$, it is possible that a few of SRAF patterns will be outside the design area when there are more than 2 via patterns.
A larger window of 2048$\times$2048 $nm^2$ will be used to capture all the SRAF patterns, which shares the same center as the design window.

\textbf{OPC, litho-simulation, and image generation.} 
We use masks and wafer patterns generated by Calibre as ground truth.
Two sets of paired data are required for training.
Mask-wafer pairs are generated to train DLS.
After that, we align design-mask-wafer data for the OPC process.
The obtained clips of size 2048$\times$2048 $nm^2$ are converted into images with 2048$\times$2048 pixels where $1nm$ represents 1 pixel.
All the 2048$\times$2048 pixels images will be centrally cropped into 1024$\times$1024 pixels images where the design window locates before training.
After training, the generated 1024 pixels images will be attached at the center of SRAF clip layer to form a 2048$\times$2048 $nm^2$ layout before testing using Calibre.
The crop-then-recover strategy saves the computational cost and improves the accuracy by focusing on the mask generation.



\subsection{Training of DLS}
\label{subsec:dls-training}
\begin{figure}[tb!]
    \includegraphics[width=0.99\linewidth]{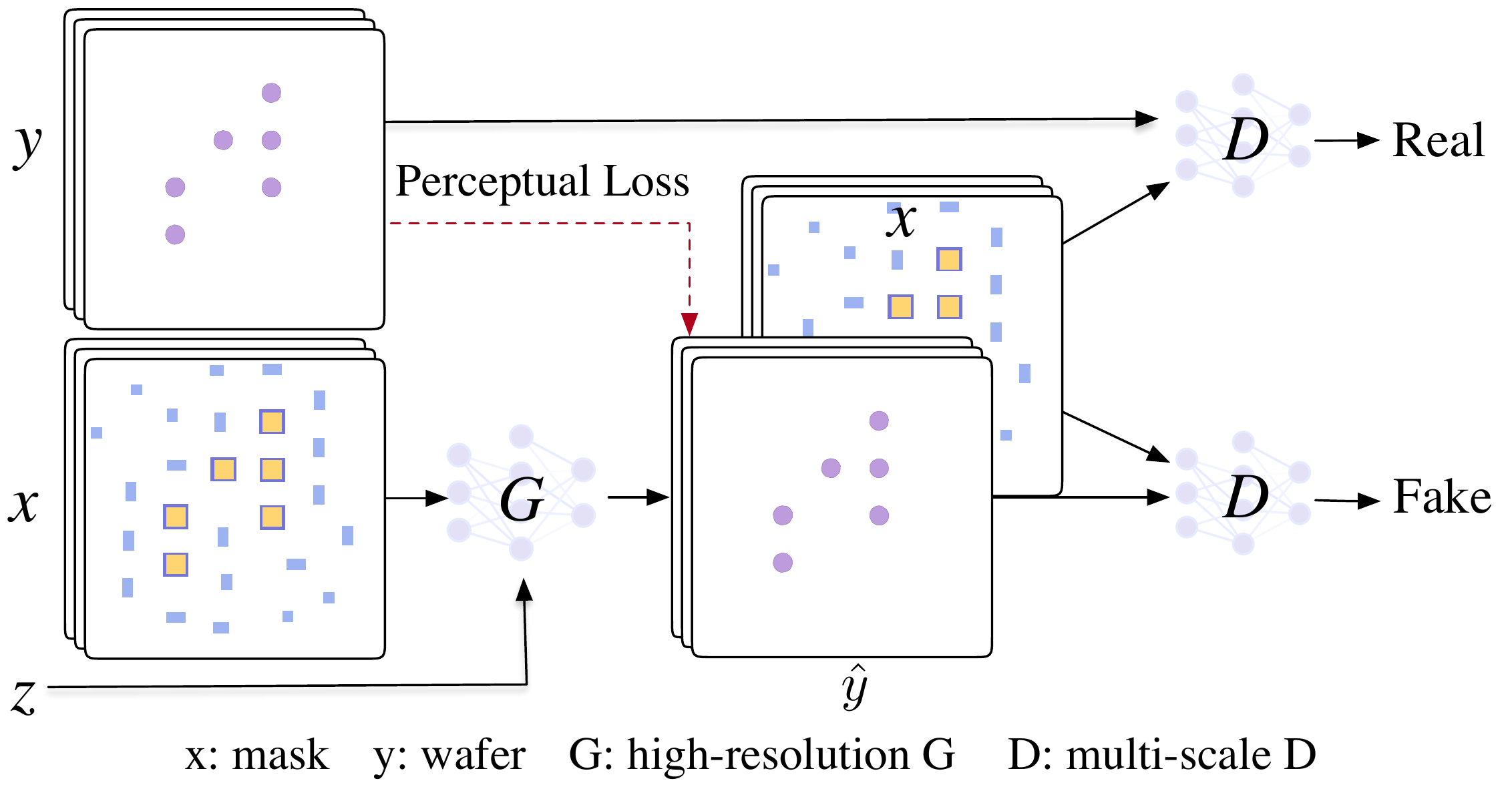}
    \centering
    \caption{The training details of DLS, where the input images are mask-wafer pairs.}
    \label{fig:cgan}
\end{figure}
\Cref{fig:cgan} shows the training process of our deep lithography simulator.
As a customized design of cGAN, DLS is trained in an alternative scheme using paired mask image $\vec{x}$ and wafer image $\vec{y}$ obtained from Mentor Calibre.
$\vec{z}$ indicates randomly initialized images.


The objectives of DLS include training the generator $G$ that produces fake wafer images $G(\vec{x}, \vec{z})$ by learning the feature distribution from $\vec{x}$--$\vec{y}$ pairs
and training the discriminators $D_1$, $D_2$ to identify the paired ($\vec{x}$, $G(\vec{x}, \vec{z})$) as fake.
This motivates the design of DLS loss function.
The first part of the loss function comes from vanilla GAN that allows the generator and the discriminator interacting with each other in an adversarial way:
\begin{equation}
		\label{eq:cgan-loss}
        \mathcal{L}_{c G A N}(G, D)
        = \mathbb{E}_{{\vec{x}}, {\vec{y}}}[\log D({\vec{x}}, {\vec{y}})] + \mathbb{E}_{{\vec{x}}, {\vec{z}}}[\log (1-D({\vec{x}}, G({\vec{x}}, {\vec{z}}))].
\end{equation}
Combined with our multi-scale discriminators described in \Cref{subsubsec:multi-d}, the \Cref{eq:cgan-loss} can be modified as:
\begin{equation}
  \label{eq:multi-d-loss-dls}
  \begin{split}
    &\sum_{k=1,2} \mathcal{L}_{cGAN}\left(G_{DLS}, D_{{DLS}_{k}}\right) =
    \sum_{k=1,2} \mathbb{E}_{{\vec{x}}, {\vec{y}}}[\log D_{DLS_{k}}({\vec{x}}, {\vec{y}})] \\
    &+ \mathbb{E}_{{\vec{x}}, {\vec{z}}}[\log (1-D_{DLS_{k}}({\vec{x}}, G_{DLS}({\vec{x}}, {\vec{z}}))],
  \end{split}
\end{equation}
where $D_{DLS_{k}}$ is the $k$th discriminator of DLS.
In DLS design, the perceptual loss is added to the objective, we denote $\hat{\vec{y}}$ as $G({\vec{x}}, {\vec{z}})$
and loss network $\Phi$ is a pre-trained VGG19 on ImageNet. The perceptual loss is given by:
\begin{align}
    \label{eq:per-loss-dls}
    \begin{split}
        \mathcal{L}_{L_P}^{G_{DLS}, \Phi}(\vec{y}, \hat{\vec{y}})=& \sum_{j=1\dots5}\mathcal{L}_{L_1}({{\phi_j}(\vec{y})}, {\phi_j(\hat{\vec{y}})})\\
        ={}& \sum_{j=1\dots5}\mathbb{E}_{{\vec{y}}, { \hat{\vec{y}} }}\left[\|{\phi_j(\vec{y})}-{\phi_j(\hat{\vec{y}})}\|_{1}\right],
    \end{split}
\end{align}
where $\phi_j$ is the feature representation on $j$-th layer of the pre-trained VGG19 $\Phi$.
By combining \Cref{eq:multi-d-loss-dls} and \Cref{eq:per-loss-dls}:
\begin{equation}
    \label{eq:DLS}
    \mathcal{L}_{DLS}= \sum_{k=1,2} \mathcal{L}_{c G A N}(G_{DLS}, D_{{DLS}_{k}})+\lambda_0 \mathcal{L}_{L_P}^{G_{DLS}, \Phi}(\vec{y}, \hat{\vec{y}}).
\end{equation}

\subsection{Training of DAMO}




\begin{figure}[tb!]
    \vspace{-.2in}
    \centering
    \subfloat[]{\includegraphics[width=.46\textwidth]{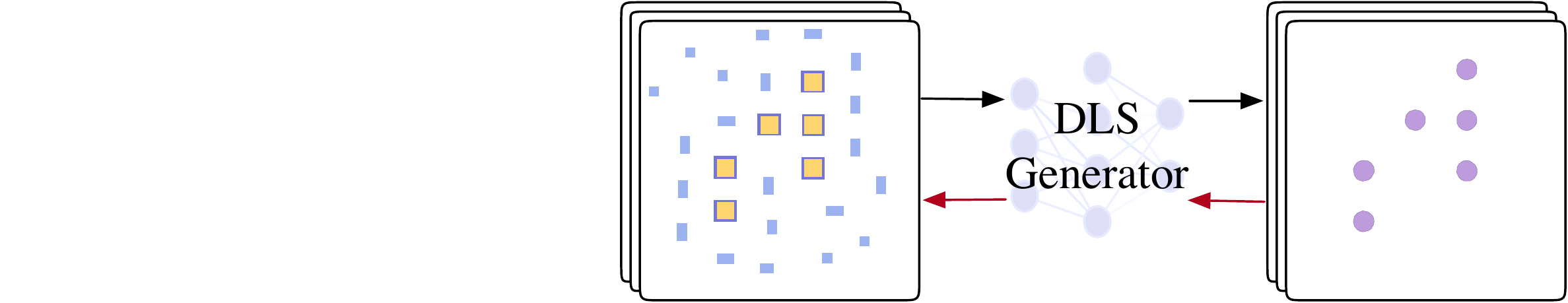} \label{fig:DLS}} \\
    \subfloat[]{\includegraphics[width=.46\textwidth]{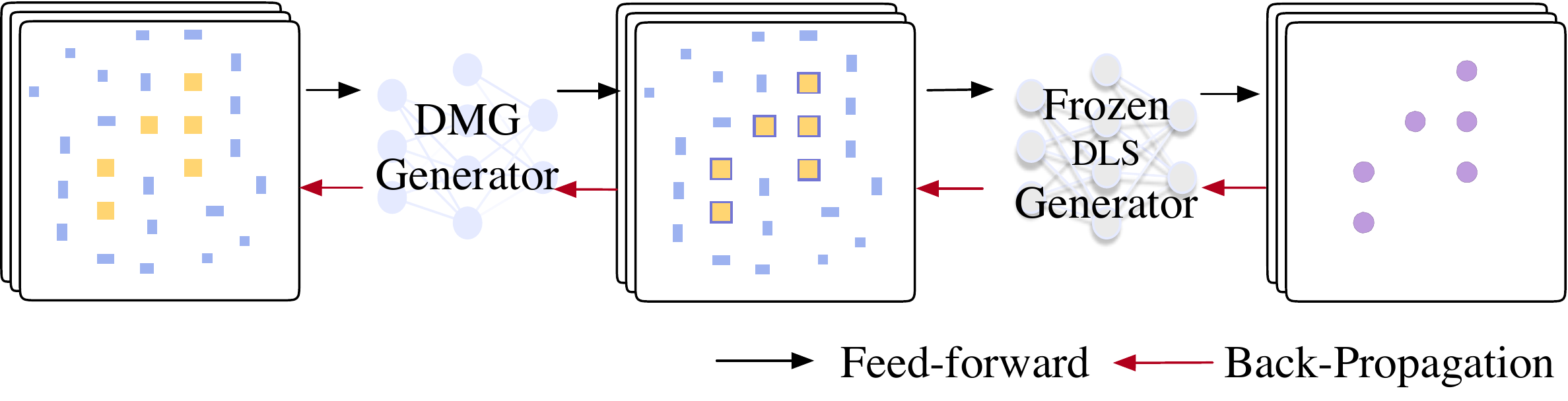} \label{fig:DAMO}}
    \caption{Overall training of DAMO: (a) Training DLS in the first stage; (b) Training DMG with fixed DLS generator in the second stage.}
    \label{fig:wholeflow} 
\end{figure}

Here we introduce the overall training procedures of the whole framework.
The first training step is illustrated in \Cref{fig:DLS}, which is focusing on DLS.
The proposed DLS is expected to predict wafer image with higher precision compared with traditional cGAN.
After the training of DLS, all parameters in its generator are frozen.

The second training step is illustrated in \Cref{fig:DAMO}, which is focusing on DMG.
DMG has the same architecture as DLS developed for DAMO training.
In this stage, training data are switched to design-mask-wafer pairs.
We use the design-mask to train DMG, obtaining an initial solution.
The objective of DMG is shown in \Cref{eq:DMG} where ${\vec{x}}$ represents the ground truth mask, $\vec{w}$ is the corresponding design, and ${\vec{z}_0}$ is the image with random values.
$G_{DMG}$, $D_{DMG}$ represents the generator and discriminator of DMG.
$\hat{{\vec{x} }}$ is the generated mask of $G_{DMG}$. Here DMG shares the same architecture as DLS, which yields a similar objective as \Cref{eq:DLS},
\begin{equation}
  \label{eq:cgan-DMG-loss}
  \begin{aligned}
      &\sum_{k=1,2} \mathcal{L}_{c G A N}(G_{DMG}, D_{DMG_{k}}) =
      \sum_{k=1,2} \mathbb{E}_{{\vec{w}}, {\vec{x}}}[\log D_{DMG_{k}}({\vec{w}}, {\vec{x}})] \\
      &+ \mathbb{E}_{{\vec{w}}, {\vec{z_0}}}[\log (1-D_{DMG_{k}}({\vec{w}}, G_{DMG}(\vec{w}, \vec{z}_0))].
  \end{aligned}
\end{equation}
\begin{equation}
    \label{eq:DMG}
    \begin{aligned}
        \mathcal{L}_{DMG}
        ={ \sum_{k=1,2}\mathcal{L}_{cGAN}(G_{DMG}, D_{DMG_{k}}})  + \lambda_1 \mathcal{L}^{G_{DMG}, \Phi}_{L_P}(\vec{x},\hat{\vec{x}}).
    \end{aligned}
\end{equation}
Then we put the solution into DLS.
RGB images instead of binary images are used because we can control the gradient of design, mask, and wafer separately, which is significant for avoiding noise points.
Separating the design, mask, and SRAF into different channels makes DAMO more stable and flexible because we can apply different weights on different channels.
After that, DLS calculates the perceptual loss between the generated wafer and the ground truth wafer. Finally, the gradient will be back-propagated to DMG to guide mask generation.
Combining \Cref{eq:DLS} with \Cref{eq:DMG}, the objective function of DAMO can be expressed as \Cref{eq:dlsopc},
\begin{equation}
    \label{eq:dlsopc}
        \mathcal{L}_{DAMO}= \mathcal{L}_{DMG} + \mathcal{L}_{DLS} + \lambda_2 \mathcal{L}_{L_1}(\hat{\vec{y}}, \vec{w}_r).
\end{equation}
We denote $\vec{w}_r$ as the via patterns (without SRAF).
The last term in \Cref{eq:dlsopc} shows the superiority of our architecture, which bridges the gap between the generated wafer ($\hat{\vec{y}}$) and target design ($\vec{w}_r$) thus optimizing the mask directly.
DAMO controls the whole flow from design to wafer while GAN-OPC relies on the conventional ILT engines.

Thanks to the guidance of DLS, our DAMO framework has a higher solution space than GAN-OPC.
The success of our approach is also verified by various experiments.
Compared to previous works, there are several advantages of DAMO:
\begin{itemize}
    \item DLS surpasses LithoGAN \cite{DFM-DAC2019-Ye} by being able to predict lithography contours of a single clip with multiple via patterns, which enables efficient training of DMG.
    \item DAMO, equipped with DCGAN-HD, can directly output manufacturing friendly masks that avoid further fine-tuning with traditional costly OPC engines.
\end{itemize}


\section{Full-chip Splitting Algorithm}
\label{sec:split_algo}

\begin{figure}[tb!]
    \vspace{-.4in}
    \centering
    \subfloat[]{ \includegraphics[width=.50\linewidth]{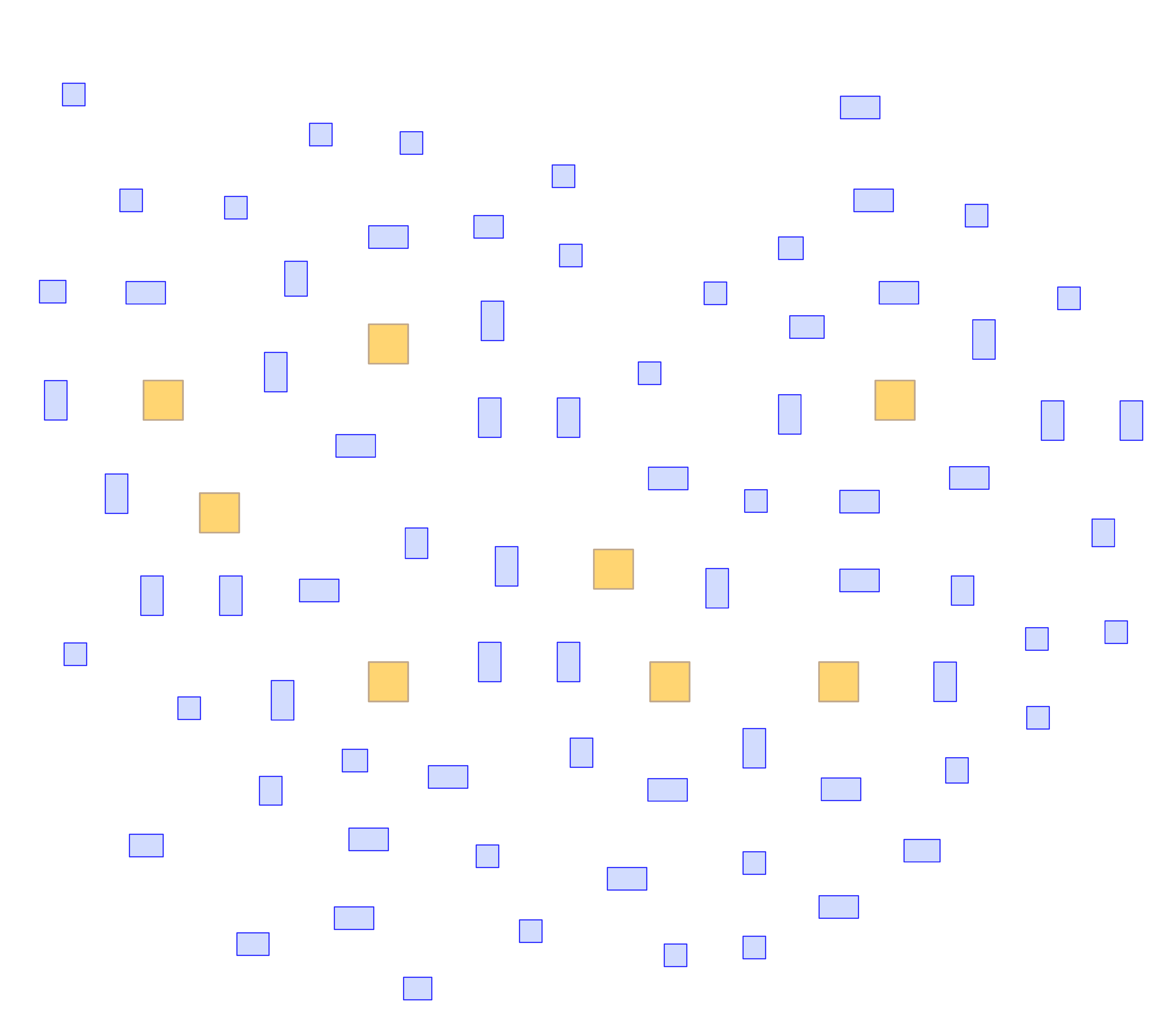} \label{fig:split_algo_via_sraf}}
    \subfloat[]{ \includegraphics[width=.50\linewidth]{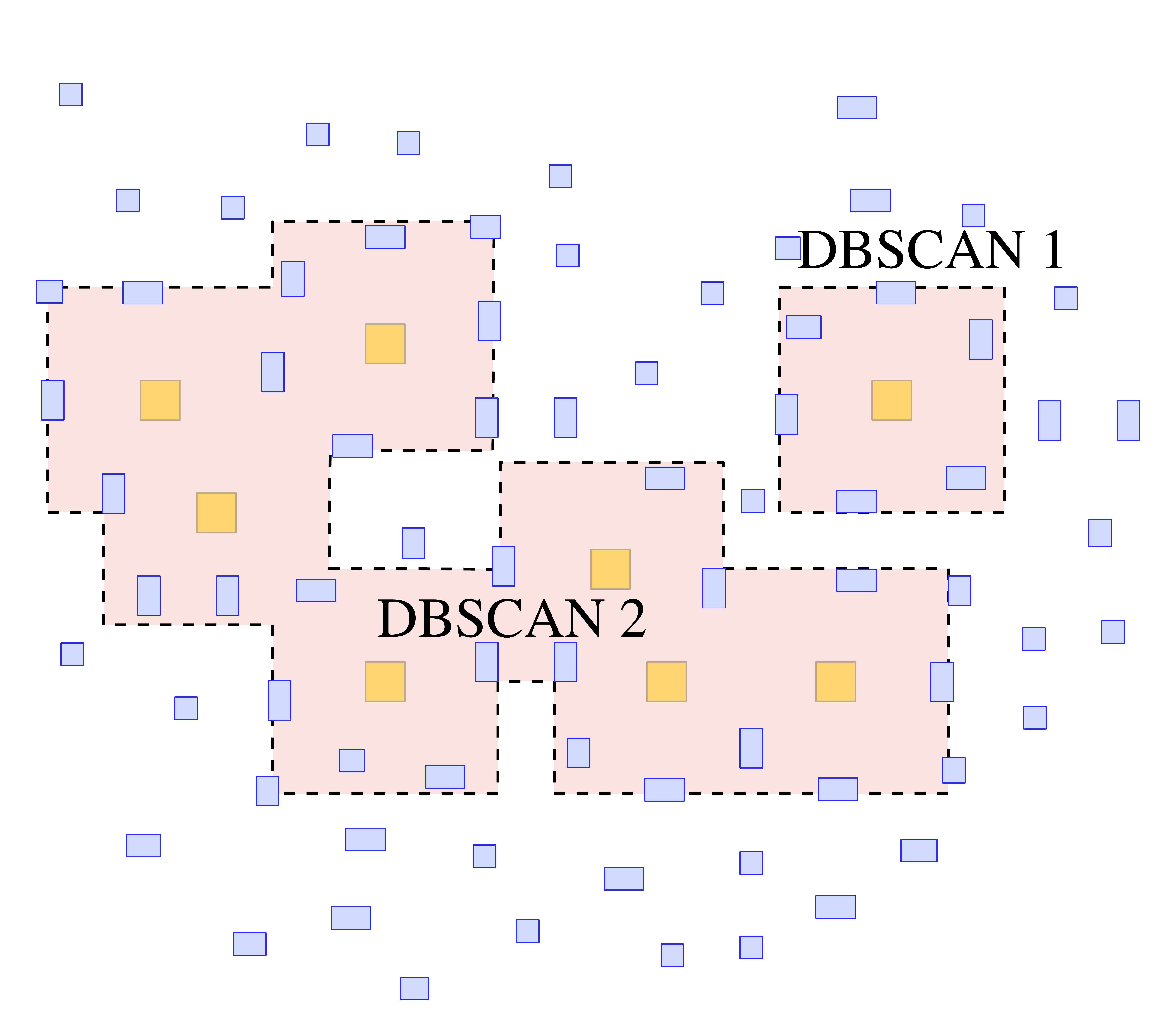} \label{fig:split_algo_dbscan}} \\
    \subfloat[]{ \includegraphics[width=.50\linewidth]{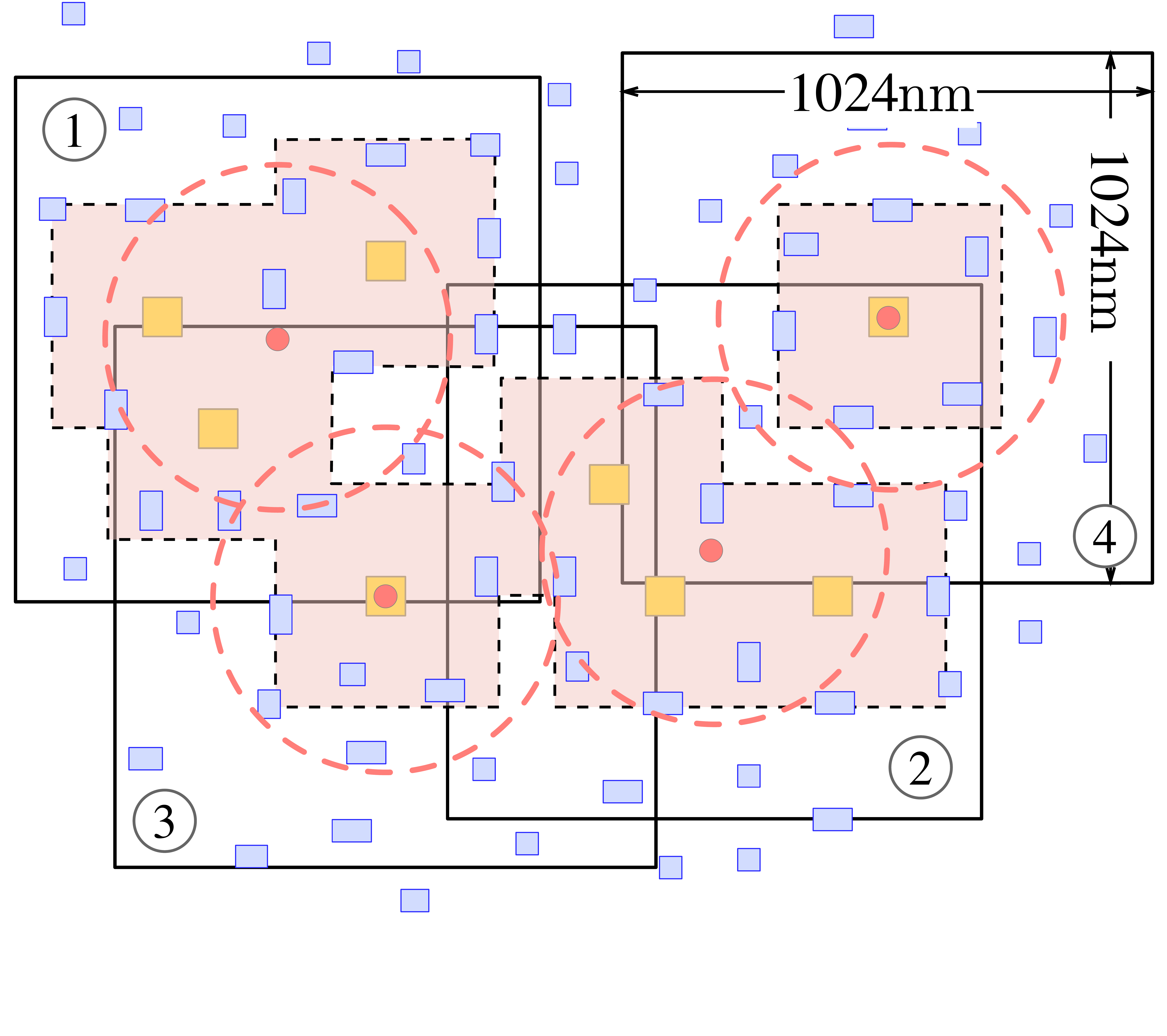} \label{fig:split_algo_kmeans}}
    \subfloat[]{ \includegraphics[width=.50\linewidth]{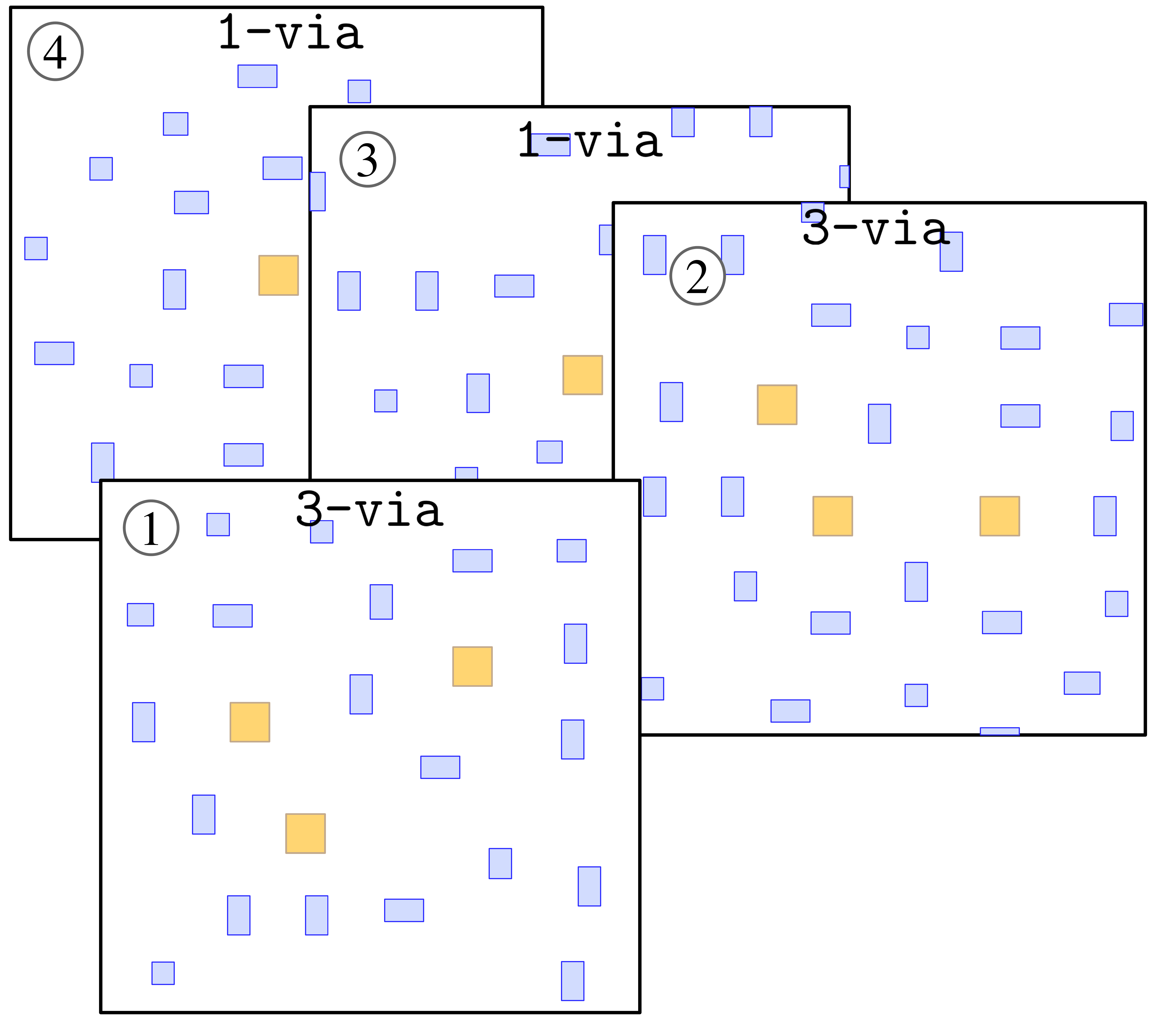} \label{fig:split_algo_windows}} \\
    \subfloat  { \includegraphics[width=.98\linewidth]{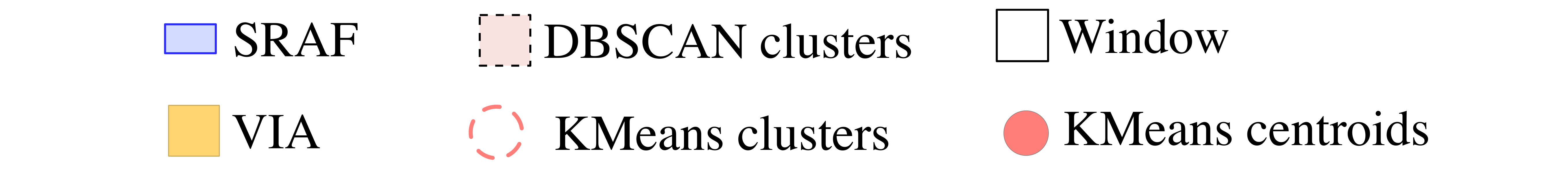}}
    \caption{
        Two-step full-chip splitting algorithm: (a) Part of full-chip; (b) Coarse step: full-chip to DBSCAN clusters;
        (c) Fine step: run KMeans++ on each DBSCAN cluster to get KMeans clusters, where each KMeans cluster belongs to a $1024 \times 1024 nm^2$ window;
        (d) The split chips.
    }
    \label{fig:split_algo}
\end{figure}


DAMO shows advantages on $1024 \times 1024 nm^2$ clips.
To further adopt DAMO on full-chip layouts, a coarse-to-fine window splitting algorithm is proposed,
in which the two-step clustering enables us to deal with full-chip industrial layouts where via patterns are distributed randomly with different local densities. A portion of one full-chip is shown in \Cref{fig:split_algo_via_sraf}.

\textbf{Coarse step: DBSCAN.} The main concept of the DBSCAN algorithm is to locate the regions of high via density that are separated from other low density regions.
Any via neighborhood within a circle of radius Eps($\epsilon$) from via $v$ will be assigned to the same cluster of $v$.
DBSCAN algorithm is used to initially detect the clusters of via patterns (lines 1--4 in \Cref{alg:split_algo}).
After the coarse step, the via patterns in a large layout will be assigned into different DBSCAN clusters, as shown in \Cref{fig:split_algo_dbscan}.

\textbf{Fine step: KMeans++.} After DBSCAN clustering, every via pattern is assigned to a coarse cluster $d$ which contains $V$ via patterns.
Then we search every coarse cluster and run KMeans++ algorithm to find the best splitting windows,
where the max number of via patterns in a window is set to $K$ (lines 5--27 in \Cref{alg:split_algo}).
Note that every KMeans cluster belongs to a $1024 \times 1024 nm^2$ window, whose center locates at the centroid of the KMeans cluster, as shown in \Cref{fig:split_algo_kmeans}.

\begin{algorithm}[h]
  \caption{Full-chip splitting algorithm.}
  \small
  \begin{algorithmic}[1]
      \Require Full-chip, DBSCAN parameter $\epsilon$;
      \Ensure Best full-chip splitting windows;
      \State $\mathcal{V} \leftarrow $ collection of all via patterns; \algorithmiccomment{DBSCAN starts.}
      \State $MinPts \leftarrow$ 1;
      \State Run DBSCAN on $\mathcal{V}$ with parameters $\epsilon$ and $MinPts$;
      \State $\mathcal{D} \leftarrow$ collection of DBSCAN clusters. \algorithmiccomment{DBSCAN ends.}
      \State $\mathcal{S} \leftarrow$ empty collection of best splitting windows;  \algorithmiccomment{KMeans++ starts.}
      \State $K \leftarrow$ max via number in a window;
      \State $H\leftarrow$ width and height of a window;
      \For{each $d \in \mathcal{D}$}
          \State $V \leftarrow$ via number in DBSCAN cluster $d$;
          \For{ $\forall k < V$}
              \State Run KMeans++ in cluster $d$ with $k$ centroids;
              \State $\mathcal{C} \leftarrow$ collection of KMeans clusters in DBSCAN cluster $d$;
              \State Create  $H \times H nm^2$ split windows centered at $k$ centroids;
              \State $BestSplitting \leftarrow$ True;
              \For{ each KMeans cluster $c \in \mathcal{C}$}
                  \State $v_c \leftarrow$ via number of KMeans cluster $c$;
                  \If { $v_c > K$ or via in $c$ is not in $k$ split windows}
                      \State $BestSplitting \leftarrow$ False;
                      \State Break;
                  \EndIf
              \EndFor
              \If {$BestSplitting$ is True}
                  \State Add the $k$ split windows to $\mathcal{S}$;
              \EndIf
          \EndFor
      \EndFor
      \State \Return collection of best splitting windows $\mathcal{S}$; \algorithmiccomment{KMeans++ ends.}
  \end{algorithmic}
  \label{alg:split_algo}
\end{algorithm}

After the coarse-to-fine clustering, the design will be split into many $1024\times1024$ $nm^2$ windows (see \Cref{fig:split_algo_windows}).
Our coarse-to-fine splitting algorithm has many advantages.
Firstly, it is extremely fast because DBSCAN only needs to scan the via patterns once and it also skips the empty areas.
Secondly, the typical window-sliding method is hard to handle overlapping situations and stitching errors.
In our algorithm, the overlapping situations and stitching errors will not occur, since every design pattern belongs to a fine cluster. 
Thirdly, because the window locates at the centroids of the clusters, the via patterns are all placed near the center of the windows, which reduces the search space of the machine learning model to a large extent, resulting in less training data and training time. 

\section{Experimental Results}

\label{sec:result}
Many experiments are carried out to evaluate our proposed framework.
Firstly, we evaluate the effectiveness of our DLS by testing the mIoU and pixAcc of generated wafer patterns.
Secondly, the superiority of our proposed DAMO is also validated by thorough experiments.
Lastly, we test our model using the full-chip layout in ISPD 2019 contest \cite{ispd2019-benchmark}, which is generated by an open-source router \cite{ROUTE-ICCAD2019-Li}.

\subsection{Dataset}
\label{subsec:dataset}

\textbf{Our training set and validation set.}
As described in \Cref{subsec:build_train}, two sets of $2048 \times 2048$ pixels RGB images are generated for training purpose: one mask-wafer paired for DLS, while another one design-mask-wafer paired for DMG.
To obtain fine-grained models, we divide our data depending on the via number with a window, and six groups marked as \texttt{1-via}, \texttt{2-via}, $\dots$ \texttt{6-via} are generated.
For instance, the \texttt{1-via} group contains all cases with only one via in a window.
Each group has 2000 training images and 500 validation images.

\begin{figure}[tb!]
  \centering
  \includegraphics[width=0.99\linewidth]{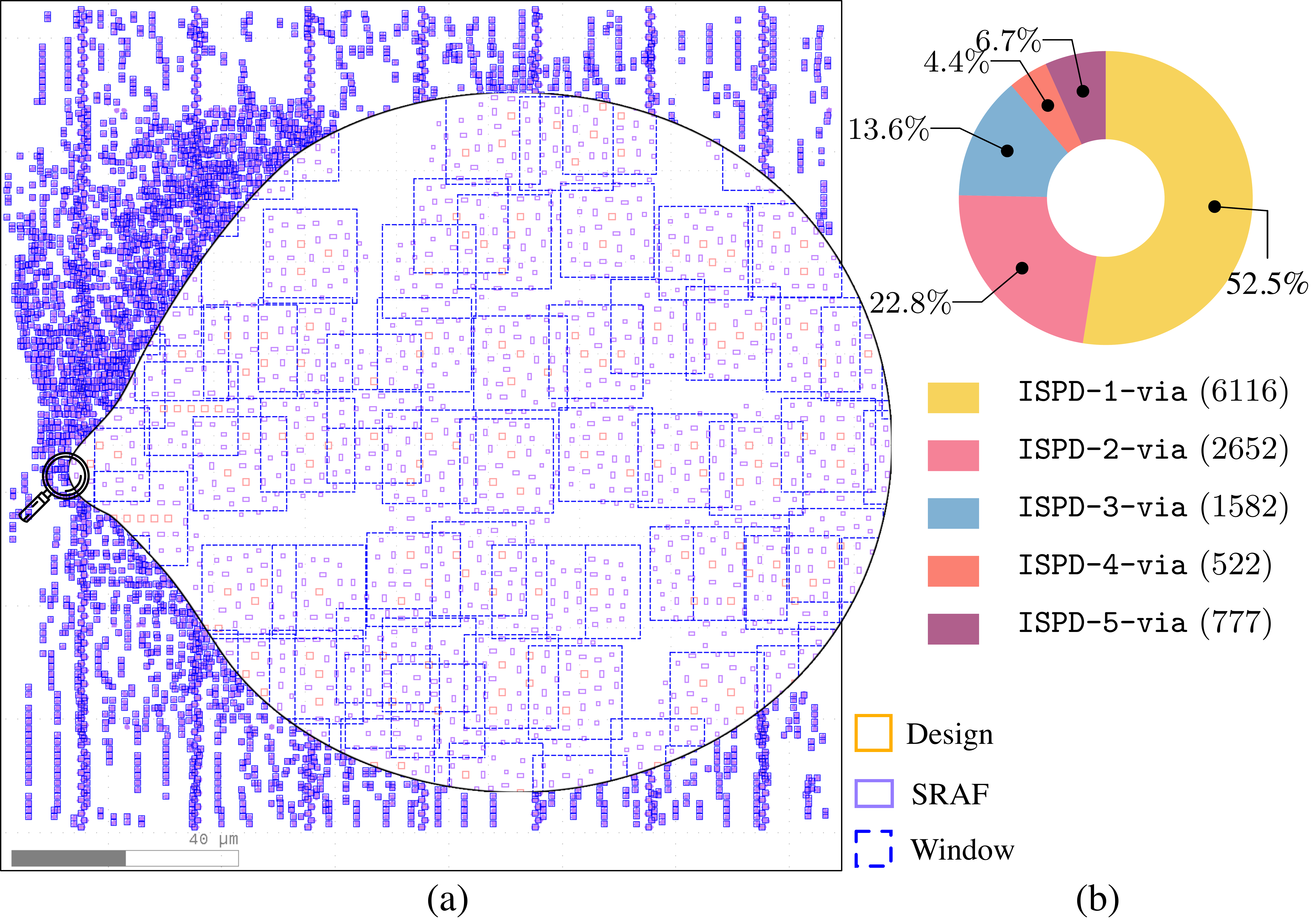} \label{fig:ispd_split_layout}
  \caption{(a) ISPD 2019 large full-chip layout and splitting windows; (b)via window distribution in $\texttt{ispd19\_test1}$\cite{ispd2019-benchmark}.}
  \label{fig:ispd_split}
\end{figure}

\noindent\textbf{ISPD 2019 large full-chip test set.}
We use another real benchmark coming from ISPD 2019 Contest on Initial Detailed Routing,
We take the layer 40 of \texttt{ispd19$\_$test1} \cite{ispd2019-benchmark} as our design layer ($100 \times 100 um$).
After the SRAF insertion, OPC, and lithography process via Calibre, we extract the design, SRAF, mask, wafer layers and merge them to be the ground truth.
Then, using our coarse-to-fine full-chip splitting algorithm,
the full-chip layout is split to lots of $1024 \times 1024 nm^2$ layout windows.
According to the design rule, we set the DBSCAN radius Eps ($\epsilon$) to be $400 nm$.
The hyper-parameters $K$ in KMeans++ fine step is set to 5, because the images containing more than 5 design patterns only account for 0.5$\%$ in the total windows.
The \texttt{ispd19$\_$test1} benchmark contains 16035 design patterns which are split to 11649 windows.
6116 split windows marked as \texttt{ISPD-1-via} has only one via in a window, accounting for $52.5\%$.
The detailed distribution of different windows is illustrated in \Cref{fig:ispd_split}.

\subsection{Implementation Details}

The proposed DAMO is implemented in Python with PyTorch library \cite{DL-NIPSW2017-PyTorch}.
Adam optimizer \cite{DL-ICLR2015-Adam} is adopted, where we set base learning rate and momentum parameters to $0.0002$ and $(0.5, 0.999)$.
In the LeakyReLU, the slope of the leak is set to 0.2 in all models.
We set the batch size to be 4, and the maximum training epoch is 100.
The weight parameters of $\lambda_0$, $\lambda_1$, and $\lambda_2$ are set to 100, 100, and 50, respectively.
After training, the generated mask layer will be converted into GDSII layout file then fed into Mentor Calibre for lithography simulation validation.
We use four Nvidia TITAN Xp GPUs for training and one for testing.
The evaluation metrics we adopt are mIoU, pixAcc, $L_2$ error, and PV Band.
Here the PV Band is calculated by Calibre.

\subsection{Effectiveness of DLS}

Before training DAMO, it is of great importance to construct a high-performance DLS.
Since our DLS model is based on the cGAN framework, we set up an ablation experiment to illustrate the advantages of our generator and discriminators.
The results shown in \Cref{tab:dcupp} is the average of 6 groups of validation set.
Firstly, cGAN (used in LithoGAN) provides a baseline mIoU of 94.16\% which is far away from practical application.
Then, UNet++ is used to replace the UNet generator in cGAN for better performance.
However, the original UNet++ is not qualified to be a generator of a cGAN and the mIoU is reduced to 93.98\% (as shown in \Cref{tab:dcupp}).

Following DCGAN, we made some amendments in UNet++ (as discussed in \Cref{subsubsec:h-res-g}) and high resolution generator is adopted in our DLS model.
After applying our high resolution generator, mIoU is improved to 97.63\%, which outperforms UNet and UNet++ generators by a large margin when using the same discriminator.
The huge gain in mIoU implies that our developed high resolution generator is a strong candidate for DLS.
Next, the newly designed multi-scale discriminators (introduced in \Cref{subsubsec:multi-d}) are used to replace the original cGAN discriminator.
Results in \Cref{tab:dcupp} show that mIoU is further boosted to 97.63\%.

Lastly, we replace the $L_1$ loss with the perceptual loss proposed in \Cref{subsubsec:per-loss} and the mIoU reaches 98.68\%.
Additionally, DLS can handle multiple vias in a single clip, which overcomes the limitation of LithoGAN \cite{DFM-DAC2019-Ye}.

\begin{table}[tb!]
    \caption{Results of DLS}
    \centering
    \renewcommand{\arraystretch}{1.08}
    \begin{tabular}{|l|l|c|c|c|}
        \hline
        Generator & Discriminator & Loss  & mIoU (\%)  & pixAcc (\%)     \\
        \hline \hline
        UNet (cGAN) & D (cGAN) & $L_1$ & 94.16  & 97.12    \\
        UNet++      & D (cGAN) & $L_1$ & 93.98  & 96.74   \\
        G (Our)    & D (cGAN) & $L_1$ & 96.23  & 97.50 \\
        G (Our)    & D (Our) & $L_1$  & 97.63  & 98.76 \\
        G (Our)    & D (Our) & Our  & \textbf{98.68}  & \textbf{99.50} \\
        \hline
    \end{tabular}
    \label{tab:dcupp}
\end{table}

\subsection{Performance of DAMO}
\begin{table*}[tbp]
  \centering
  \caption{Comparison with State-of-the-art on validation set}
  \label{tab:dls_opc}
  \setlength{\tabcolsep}{1.8pt}
  \renewcommand{\arraystretch}{1.08}
  \begin{tabular}{|cc|ccc|ccc|ccc|}
      \hline
      \multirow{2}{*}{Bench} & \multirow{2}{*}{case\#} & \multicolumn{3}{c|}{GAN-OPC} & \multicolumn{3}{c|}{Calibre} &\multicolumn{3}{c|}{DAMO} \\
      & &$L_2$ ($nm^2$) &PV Band ($nm^2$) &runtime (s) &$L_2$ ($nm^2$) &PV Band ($nm^2$) &runtime (s)    &$L_2$ ($nm^2$) &PV Band ($nm^2$) &runtime (s) \\ \hline \hline
      \texttt{1-via}& 500 &1464 &3064      &321      &1084 &2918 &1417          &\textbf{1080} &\textbf{2917}     &\textbf{284}  \\
      \texttt{2-via}& 500 &4447 &6964      &336      &2161 &5595 &1406         &\textbf{2129} &\textbf{5576}      &\textbf{281} \\
      \texttt{3-via}& 500 &8171 &11426     &317      &3350 &8286 &1435         &\textbf{3244} &\textbf{8271}      &\textbf{285} \\
      \texttt{4-via}& 500 &11659 &14958    &327      &4331 &10975 &1477        &\textbf{4263} &\textbf{10946}      &\textbf{291} \\
      \texttt{5-via}& 500 &15773 &18976    &318      &5410 &13663 &1423        &\textbf{5396} &\textbf{13640}    &\textbf{279}    \\
      \texttt{6-via}& 500 &18904 &22371    &320      &6647 &15572 &1419        &\textbf{5981} &\textbf{15543}     &\textbf{284}  \\ \hline
      \multicolumn{2}{|c|}{Average}  &10069 &12960  &323             &3831 &9502 &1430          &\textbf{3682} &\textbf{9482}       &\textbf{284} \\
      \multicolumn{2}{|c|}{Ratio}  &2.735 &1.367 &1.138            &1.040 &1.002  &4.427   &\textbf{1.00} &\textbf{1.00}  &\textbf{1.00} \\ \hline
  \end{tabular}
\end{table*}

\begin{table*}[tbp]
  \centering
  \caption{Comparison on ISPD 2019 full-chip splitting windows}
  \label{tab:damo_ispd}
  \setlength{\tabcolsep}{1.8pt}
  \renewcommand{\arraystretch}{1.08}
  \begin{tabular}{|cc|ccc|ccc|ccc|}
      \hline
      \multirow{2}{*}{Bench} & \multirow{2}{*}{case\#} & \multicolumn{3}{c|}{GAN-OPC} & \multicolumn{3}{c|}{Calibre} &\multicolumn{3}{c|}{DAMO} \\
      & &$L_2$ ($nm^2$) &PV Band ($nm^2$) &runtime (s) &$L_2$ ($nm^2$) &PV Band ($nm^2$) &runtime (s)    &$L_2$ ($nm^2$) &PV Band ($nm^2$) &runtime (s) \\ \hline \hline
      \texttt{ISPD-1-via} &6116    &2367  &3492                      & 3963   &1073 &2857 &18959            &\textbf{1056} &\textbf{2848}          &\textbf{        3669     }    \\
      \texttt{ISPD-2-via} &2652    &5412  &7126                      & 1742    &2232 &5670 &7537           &\textbf{2172} &\textbf{5654}           &\textbf{        1591     }    \\
      \texttt{ISPD-3-via} &1582    &8792  &13047                     & 1021   &3602 &8276 &4494           &\textbf{3196} &\textbf{8127}            &\textbf{        949      }    \\
      \texttt{ISPD-4-via} &522     &12395 &15015                    & 341    &4395 &11051 &1692          &\textbf{4361 } &\textbf{10987}          &\textbf{        313      }     \\
      \texttt{ISPD-5-via} &777     &16526 &19147                    & 495    &5526 &12305 &2537          &\textbf{4542} &\textbf{12251}           &\textbf{        466      }     \\ \hline
      \multicolumn{2}{|c|}{Average} &9098  &11565 &1512  &3365 &8031 &7043           &\textbf{3065} &\textbf{7973}               &\textbf{ 1397 }\\
      \multicolumn{2}{|c|}{Ratio}   &2.968 &1.451  &1.082   &1.098 &1.007  &5.041     &\textbf{1.00} &\textbf{1.00}  &\textbf{1.00} \\
      \hline
  \end{tabular}
\end{table*}

\begin{figure}[tb!]
    \centering
    \subfloat[]{ \includegraphics[width=.18\linewidth]{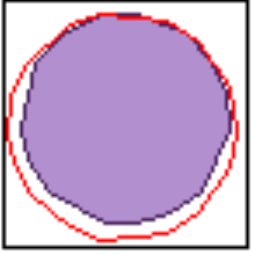} \label{fig:visual-via-1}}
    \subfloat[]{ \includegraphics[width=.18\linewidth]{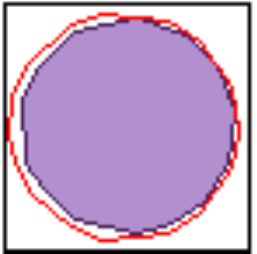} \label{fig:visual-via-2}}
    \subfloat[]{ \includegraphics[width=.18\linewidth]{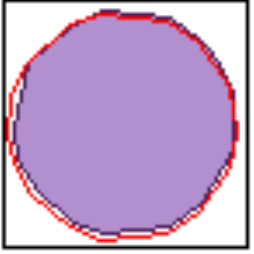} \label{fig:visual-via-3}}
    \subfloat[]{ \includegraphics[width=.18\linewidth]{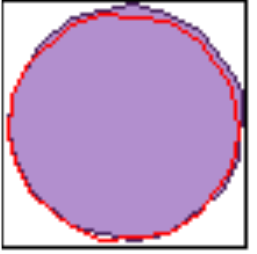} \label{fig:visual-via-4}}
    \subfloat[]{ \includegraphics[width=.18\linewidth]{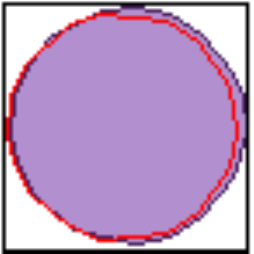} \label{fig:visual-via-5}}
    \caption{
        Visualization of DAMO model advancement on via layer:
        (a) Epoch 20; (b) Epoch 40; (c) Epoch 60; (d) Epoch 80; (e) Epoch 100. 
    }
    \label{fig:visual-via}
\end{figure}

We test DAMO on the six groups of validation sets to verify the performance.
Every generated mask will be pushed into Calibre for lithography simulation.
After that, we apply $L_2$ and PV Band measurements to test the performance of different mask optimization methods.
Note that since GAN-OPC fails to train on high-resolution input, the 1024$\times$1024 input images are downsampled to 256$\times$256 pixels to train the model.
After the inference process, the results are upsampled to the original size for $L_2$ and PV Band testing.
\Cref{tab:dls_opc} shows that on the validation set, DAMO has $2.7\times$ less $L_2$ error and $1.3\times$ less PV Band compared with GAN-OPC.
In addition, DAMO outperforms Calibre in both $L_2$ and PV Band metrics, meanwhile achieving $4\times$ speed-up.
The $L_2$, the PV Band, and the runtime performance of DAMO are better than Calibre and GAN-OPC in all cases, which demonstrates that the stability of DAMO can be guaranteed.

The mask optimization process of DAMO is visualized in \Cref{fig:visual-via}.
All the wafer images are generated using Calibre lithography simulation.
The red contours represent wafer patterns on masks produced by Calibre while the purple wafers are on masks generated by DAMO.
We sample DAMO results after 20/40/60/80/100 training epochs for the illustration.
Initially, the wafer patterns of DAMO have lower quality compared with Calibre (as shown in \Cref{fig:visual-via-1} and \Cref{fig:visual-via-2}).
Along with the increase of training epochs, the results of DAMO and Calibre are getting closer (\Cref{fig:visual-via-3}).
\Cref{fig:visual-via-4} and \Cref{fig:visual-via-5} show that the performance of DAMO surpasses Calibre after iterative optimization.

\subsection{Results on ISPD 2019 Full-chip Layout}
\label{subsec:ispd19_res}
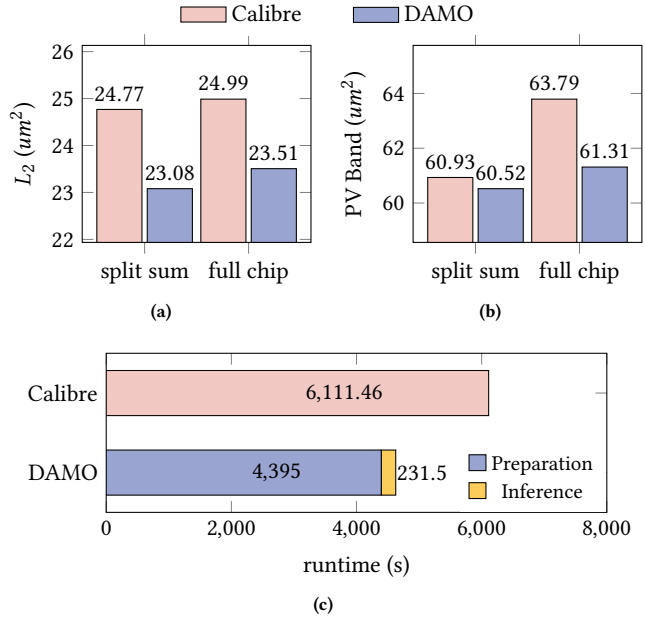
\begin{figure}[tb!]
    \vspace{-.2in}
    \centering
    \subfloat []{
\pgfplotsset{
    width =0.55\linewidth,
    height=0.5\linewidth
}

\begin{tikzpicture}
  \begin{axis}[
      ybar,
      enlargelimits=0.6,
      bar width=0.6cm,
      legend style={at={(0.5, 1)},
      anchor=north,legend columns=-1},
      ylabel={$L_2$ ($um^2$)},
      symbolic x coords={split sum, full chip},
      xtick=data,
      ylabel near ticks,
      xlabel near ticks,
      nodes near coords,
      nodes near coords align={vertical},
      legend style={
        draw=none,
        at={(1.02,1.05)},
        anchor=south east,
        legend columns=-1,
    }
      ]
  \addplot [ybar, fill=myred, draw=black,area legend] coordinates {(split sum, 24.7680) (full chip, 24.9874) };
  \addplot [ybar, fill=myblue, draw=black,area legend] coordinates {(split sum, 23.0803) (full chip, 23.5065) };
  \legend{ Calibre },
  \end{axis}  
\end{tikzpicture}      \label{fig:l2_compare}}
    \subfloat []{
\pgfplotsset{
    width =0.55\linewidth,
    height=0.5\linewidth
}
\begin{tikzpicture}
  \begin{axis}[
      ybar,
      enlargelimits=0.6,
      bar width=0.6cm,
      legend style={at={(0.5, 1)},
      anchor=north,legend columns=-1},
      ylabel={PV Band ($um^2$)},
      ylabel near ticks,
      symbolic x coords={split sum,full chip},
      xtick=data,
      nodes near coords,
      nodes near coords align={vertical},
      legend style={
          draw=none,
          at={(-0.3,1.05)},
          anchor=south west,
          legend columns=-1,
      }
  ]
  \addplot [ybar, fill=myred, draw=black,area legend] coordinates {(split sum, 60.9325) (full chip, 63.7928)};
  \addplot [ybar, fill=myblue, draw=black,area legend] coordinates {(split sum, 60.5239) (full chip, 61.3132)};
  \legend{,DAMO}
  \end{axis}
  \end{tikzpicture}     \label{fig:pvb_compare}} \\
    \subfloat []{\definecolor{myyellow}{RGB}{251, 205, 91}
\definecolor{myred}{RGB}{242, 200, 195}
\definecolor{myblue}{RGB}{154, 164, 208}

\pgfplotsset{
    width =0.98\linewidth,
    height=0.44\linewidth
}

\begin{tikzpicture}
\begin{axis}[
xbar stacked, xmin=0,
bar width=0.6cm,
yticklabels={DAMO, Calibre}, ytick={1,3},
ymax=4,
ymin=0,
xmin=0,
xmax=8000,
xlabel={runtime (s)},
nodes near coords,
nodes near coords align=horizontal, 
legend style={at={(1.0, 0.0), legend font = \small},
draw=none,anchor=south east,legend columns=1},
]
\addplot [fill=myred] coordinates {
    (0,1)
    (6111.46,3)};
\addplot [fill=myblue] coordinates {
    (4395,1)
    (0,3)};
\addplot [fill=myyellow] coordinates {
    (231.5,1)
    (0,3)};
\legend{,Preparation,Inference}
\end{axis}
\end{tikzpicture} \label{fig:runtime_compare}}
    \caption{Comparison with Calibre on ISPD 2019 full-chip layout in terms of (a) L2; (b) PV Band; (c) runtime.}
    \label{fig:dlsopc_ispd_large}
\end{figure}

For ISPD 2019 large full-chip layout, the experiment has two stages.
In the first stage, we test DAMO on the 11649 split windows, as listed in \Cref{tab:damo_ispd}.
We compare GAN-OPC, Calibre, and DAMO under metrics of L2, PV Band, and runtime.
DAMO shows better performance against Calibre and GAN-OPC, on all metrics of $L_2$, PV Band, and runtime.


In the second stage, we recover all the split windows into the original 100$\times$100 $um^2$ large full-chip layout with DAMO generated masks.
Still, we use Calibre to test the $L_2$ error and PV Band of the large layout results.
\Cref{fig:dlsopc_ispd_large} shows the sum of L2 error and PV band on split windows are very close to the results of full-chip layouts owing to our efficient splitting algorithm.
As shown in \Cref{fig:l2_compare} and \Cref{fig:pvb_compare}, DAMO still has better performance than Calibre.
For the runtime of the large full-chip layout (see \Cref{fig:runtime_compare}), we separate runtime of DAMO to preparation time (4395s) and inference time (231.5s).
The inference time takes only 5$\%$ of the total by parallel using four GPUs.
Preparation includes the full-chip splitting, split layouts to images, generated images to layouts, and the split windows to full-chip recovering.
All these preparation processes are running on a single CPU, which means the preparation time can be easily reduced when using multi CPUs in parallel.

\section{Conclusion}
\label{sec:conclu}
In this paper, we present DAMO, an end-to-end framework targeting full-chip mask optimization with high-resolution generative machine learning models.
The framework comes with DLS that offers precise lithography prediction benefiting from the proposed DCGAN-HD.
The high-quality DLS also enables efficient training of DMG which hence promises to generate manufacturing friendly masks without further costly fine-tuning.
The advantage of the proposed framework over the representative industrial and academic state-of-the-art demonstrates the possibility of deep neural networks as an alternative solution to many layout and mask optimization problems.
Our future research includes the deployment of the framework to more complicated designs (such as metal layers) and the transfer-ability as technology node advances.

\section{Acknowledgment}

This work is partially supported by
The Research Grants Council of Hong Kong SAR (No.~CUHK14209420).

\clearpage
{
\bibliographystyle{IEEEtran}
\bibliography{./ref/Top,./ref/HSD,./ref/DFM,./ref/DL,./ref/PD,./ref/Software}
}


\end{document}